\documentclass{iau}
\usepackage{graphicx,natbib}

\def\R23{\mbox{$\rm R_{23}$}}

\def\Hb{\mbox{${\rm H}{\beta}$}}
\def\Ha{\mbox{${\rm H}{\alpha}$}}
\def\OIIIa{\mbox{${\rm [O\,III]\,}{\lambda\,5007}$}}

\def\OII{\mbox{${\rm [O\,II]\,}{\lambda\,3727}$}}

\def\NII{\mbox{${\rm [N\,II]\,}{\lambda\,6584}$}}

\title{Oxygen abundances of zCOSMOS galaxies at $z \sim 1.4$ based on five lines and implications for the fundamental metallicity relation}

\author[Maier et al.]{Christian Maier$^1$, Simon J. Lilly$^2$, Bodo L. Ziegler$^1$, zCOSMOS team}

\affiliation{$^1$Department for Astrophysics, University of Vienna,\\ T\"urkenschanzstr. 17, 1180 Vienna, Austria \\
email: {\tt christian.maier@univie.ac.at} \\
$^2$Institute of Astronomy, ETH Zurich, 8093 Zurich, Switzerland}

\pubyear{2014}
\volume{309}
\jname{Galaxies in 3D across the Universe}
\editors{B. L. Ziegler, F. Combes, H. Dannerbauer, M. Verdugo, eds.}

\begin{document}

\maketitle

\begin{abstract}
A relation between the stellar mass $M$ and  the gas-phase metallicity $Z$ of galaxies, the MZR, is observed up to higher redshifts. It is a matter of debate, however, if the SFR is a second parameter in the MZR.
%
To explore this issue at $z>1$, we used VLT-SINFONI  near-infrared (NIR) spectroscopy of eight zCOSMOS galaxies at $1.3<z<1.4$ to measure the strengths of  four emission lines: \Hb, \OIIIa, \Ha, and \NII, additional to \OII\, measured from  VIMOS.  We derive reliable O/H metallicities based on five lines, and also SFRs from extinction corrected \Ha\, measurements.
%
We find that the MZR of these star-forming galaxies at $z \approx 1.4$ is lower than the local SDSS MZR by a factor of three to five, a larger change than reported in the literature using [NII]/\Ha-based metallicities from individual and stacked spectra. Correcting N2-based O/Hs using recent results by \citet{newman14},
also the larger FMOS sample at $z \sim 1.4$ of \citet{zahid14} shows a similar evolution of the MZR  like the zCOSMOS objects.
These observations seem also in agreement with a non-evolving FMR using the physically motivated formulation of the FMR from \citet{lilly13}.
\end{abstract}

\firstsection
\section{Introduction}
In the local universe, the metallicity at a given mass also depends on the SFR  of the galaxy
\citep[e.g.,][]{mannu10},
i.e., the SFR appears to be a ``second-parameter'' in the mass-metallicity relation (MZR). 
\citet{mannu10} claimed that this local $Z(M,SFR)$  is also applicable to higher redshift galaxies, and coined the phrase fundamental metallicity relation (FMR) to denote this \emph{epoch-invariant} $Z(M,SFR)$ relation.
\citet{lilly13} showed that a $Z(M,SFR)$ relation is a natural outcome of a simple model of galaxies in which the SFR is regulated by the mass of gas present in a galaxy.

Most studies of the Z(M,SFR) at $z \sim 1.4$ were based on small samples or samples with limited spectroscopic
 information, producing contradictory results. 
Except for the $z \sim 1.4$ study  of \citet{maier06},
the derived O/H metallicities in the literature at $z \sim 1.4$ have been based just on [NII]/H$\alpha$, via the N2-method \citep{petpag04}.
However, \citet{newman14} found that the MZR at high $z$ determined using  the N2-method  might be  up to a factor of three times too high in terms of metallicity, when the effects of both photoionization and shocks are not taken into consideration.

\section{Results}

We have carried out NIR follow-up spectroscopy with  VLT-SINFONI of eight massive star-forming galaxies at $1.3<z<1.4$, selected from the zCOSMOS-bright sample \citep{lilly09} based on their  \OII\, emission line observed with VIMOS.  
Observations with SINFONI in two bands (J to observe the \Hb\, and \OIIIa, and H to observe \Ha\, and \NII) were performed between November 2013 and January 2014.
SFRs, masses and O/H metallicities were measured as described in \citet{maier14}.


\begin{figure}
\centering
\includegraphics[width=0.5\columnwidth,angle=270]{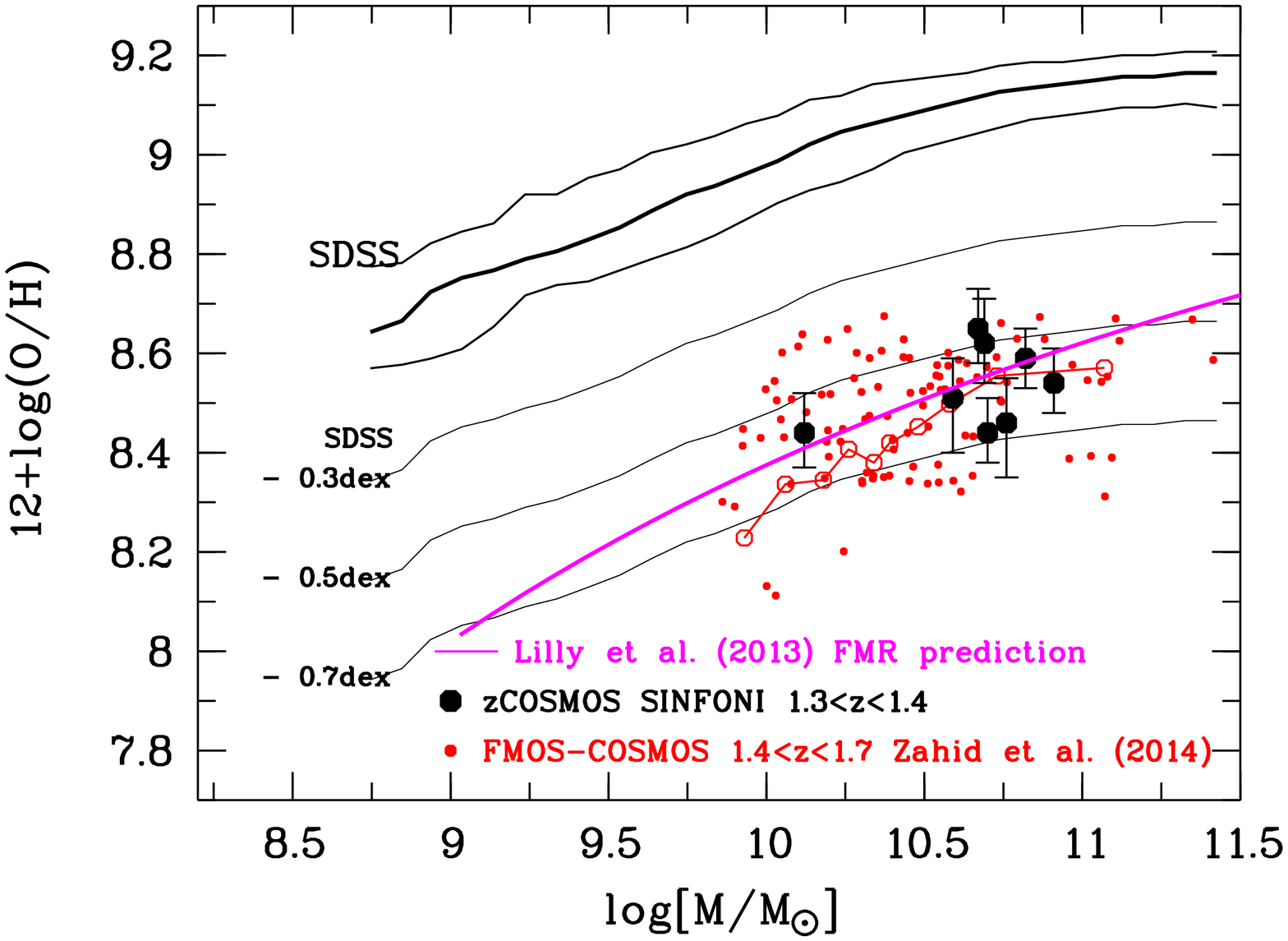} 
\caption{The median local MZR and 1$\sigma$ values of local SDSS galaxies from \citet{trem04} are shown by the three upper black lines, while the three thinner lower black lines show the median SDSS MZR shifted downward by 0.3, 0.5 and 0.7 dex, respectively. 
The  \citet{zahid14} N2-based O/Hs are corrected for the N2-calibration issues discussed in \citet{newman14}, and shown as red dots for the individual measurements, and as open red circles for the O/Hs derived from the stacked spectra.
The observed MZR  at $z \approx 1.4$ is lower than the local SDSS MZR by a factor of three to five, and is also in agreement with the FMR prediction of \citet{lilly13}.
}\label{MZRz14}
\end{figure}


In Fig.\,\ref{MZRz14} we compare the MZR of  galaxies at $1.3<z<1.7$ with the relation of SDSS. Stellar masses are converted to or computed with a \citet{salp55} IMF, O/Hs are computed or converted to the \citet{kewdop02} calibration.
The \citet{zahid14} N2-based O/Hs were converted to the  \citet{kewdop02} calibration and then systematically corrected to 0.4\,dex lower metallicities, in agreement with recent results of \citet{newman14} who found that the MZR at $z>1$ determined using  the N2-method might be  $2-3$ times too high in terms of metallicity.
We also overplot (as a magenta line) the \citet{lilly13} prediction of a non-evolving Z(M,SFR), the case represented by the dashed lines in Fig.\,7 of \citet{lilly13}.
The $z \sim 1.4$ observational data shown in Fig.\,\ref{MZRz14} are in agreement with this FMR prediction.




\end{document}